\documentclass[final, 11pt]{elsarticle}

\usepackage{natbib}%
\usepackage[figuresright]{rotating}

\usepackage{url}
\usepackage{amsmath,amsthm}

\newtheorem{lemma}{Lemma}

\theoremstyle{definition}

\theoremstyle{remark}

\usepackage[top=3.5cm,bottom=3.5cm,left=3.5cm,right=3.5cm]{geometry}

\biboptions{authoryear}

\usepackage{subcaption}
\usepackage{algorithm}
\usepackage{algpseudocode}
\usepackage{eurosym}
\usepackage{booktabs}

\usepackage{tikz}
\usetikzlibrary{arrows,automata,backgrounds,fit,patterns,petri,shadows,shapes}
\pgfdeclarepatternformonly{stripes}
{\pgfpointorigin}{\pgfpoint{0.25cm}{0.25cm}}
{\pgfpoint{0.25cm}{0.25cm}}
{
    \pgfpathmoveto{\pgfpoint{0cm}{0cm}}
    \pgfpathlineto{\pgfpoint{0.25cm}{0.25cm}}
    \pgfpathlineto{\pgfpoint{0.25cm}{0.12cm}}
    \pgfpathlineto{\pgfpoint{0.12cm}{0cm}}
    \pgfpathclose%
    \pgfusepath{fill}
    \pgfpathmoveto{\pgfpoint{0cm}{0.12cm}}
    \pgfpathlineto{\pgfpoint{0cm}{0.25cm}}
    \pgfpathlineto{\pgfpoint{0.12cm}{0.25cm}}
    \pgfpathclose%
    \pgfusepath{fill}
}
\tikzstyle{decisionA} = [regular polygon, regular polygon sides = 4, thick, minimum size = 1.25cm, inner sep = 0.1pt, draw = black, fill = gray!40]
\tikzstyle{decisionC} = [regular polygon, regular polygon sides = 4, thick, minimum size = 1.25cm, inner sep = 0.1pt, draw = black]
\tikzstyle{utilityA} = [regular polygon, regular polygon sides = 6, thick, minimum size = 1cm, inner sep = 0.1pt, draw = black, fill = gray!40]
\tikzstyle{utilityC} = [regular polygon, regular polygon sides = 6, thick, minimum size = 1cm, inner sep = 0.1pt, draw = black]
\tikzstyle{chanceA} = [circle, thick, minimum size = 1cm, inner sep = 0.1pt, draw = black, fill = gray!40]
\tikzstyle{chanceC} = [circle, thick, minimum size = 1cm, inner sep = 0.1pt, draw = black]
\tikzstyle{chance} = [circle, thick, minimum size = 1cm, inner sep = 0.1pt, draw = black, pattern = stripes, pattern color = gray!40]
\tikzstyle{empty} = [circle, line width = 0pt, minimum size = 1cm, inner sep = 0.1pt]

\DeclareMathOperator*{\argmax}{arg\,max}

\newcommand{\dd}{\mathop{}\! \mathrm{d}}
\DeclareSymbolFont{largesymbolsA}{U}{txexa}{m}{n}
\DeclareMathSymbol{\varprod}{\mathop}{largesymbolsA}{16}

\makeatletter
\def\ps@pprintTitle{%
   \let\@oddhead\@empty
   \let\@evenhead\@empty
   \let\@oddfoot\@empty
   \let\@evenfoot\@empty
}
\makeatother

\begin{document}

\title{Personalized Pricing Decisions Through Adversarial Risk Analysis}

\author[my1address]{Daniel García Rasines\fnref{equalcontribution}}
\author[my1address]{Roi Naveiro\fnref{equalcontribution}} 
\author[my2address]{David Ríos Insua\fnref{equalcontribution}}
\author[my3address]{Simón Rodríguez Santana\fnref{equalcontribution}}


\address[my1address]{CUNEF Universidad, Calle Almansa 101, Madrid, 28040, Spain}

\address[my2address]{Institute of Mathematical Sciences, Calle Nicolás Cabrera, 13-15, Madrid, 28049, Spain}

\address[my3address]{ICAI School of Engineering, Comillas University, Calle de Alberto Aguilera, 25 Madrid, 28015, Spain}

\fntext[equalcontribution]{All authors contributed equally, listed in alphabetical order.}


\begin{abstract}
Pricing decisions stand out as one of the most critical tasks a company faces, particularly in today’s digital economy. 
\textcolor{black}{ As with other business decision-making problems}, pricing unfolds in a highly competitive and uncertain environment. Traditional analyses in this area have heavily relied on game theory and its variants. However, an important drawback of these approaches is their reliance on common knowledge assumptions, which are hardly tenable in competitive business domains. This paper introduces an innovative personalized pricing framework designed to assist decision-makers in undertaking pricing decisions amidst competition, considering both buyer's and competitors' preferences. Our approach (i) establishes a coherent framework for modeling competition mitigating common knowledge assumptions; (ii) proposes a principled method to forecast competitors' pricing and customers' purchasing decisions, acknowledging 
  \textcolor{black}{major} business uncertainties; and, (iii) encourages structured thinking about the competitors' problems, thus enriching the solution process. \textcolor{black}{ To illustrate these properties}, in addition to a general pricing template, we outline two specifications -- one from the retail domain and a more intricate one from the pension fund domain.
\end{abstract}


\maketitle

\section{Introduction}

Supporting pricing decisions is one of the most critical tasks a company faces. Business magnate Warren Buffet referred to pricing power as ``the single most important decision in evaluating a business"~\citep{EconTimes2011}. Even though it has been traditionally acknowledged as a key marketing element \citep{Morris1987} pricing is particularly important in today’s digital economy, with many companies having access to large quantities of pricing-related data and high computing power that allows them to make better-informed decisions~\citep{OECD}. This is especially important in times of austerity, with companies seeing their sales curtailed and facing costs that can hardly be reduced~\citep{Economist2013}. Ultimately, smart pricing stands as a major 
strategic business tool. 

{\color{black} Recently, pricing algorithms based on machine learning (ML) have replaced more traditional, theory-based approaches in some sectors.}
These algorithms allow companies to leverage \textcolor{black}{ more information, when available, and react dynamically } according to changes in demand and competitors' movements. Indeed, many companies employ real-time pricing, whereby prices are automatically adjusted whenever market conditions change~\citep{Rana2014, Chen2016},
and customer features are utilized for personalized pricing purposes~\citep{Choudhary2005}. This is part of a broader trend where automated approaches \textcolor{black}{ to decision support are gaining widespread adoption across diverse business landscapes \citep{gupta2022artificial}. 
Yet the implications of the extensive use of pricing algorithms are 
to some degree still unknown}. {\color{black} In some contexts, concerns have been raised about the explainability, interpretability, and safety of the usage of some ML models in sensitive domains \citep{bibal2021legal}, which has led to legal frameworks that should be taken into account depending on the context \citep{GDPR,SR117}}. For our particular case, expectations suggest that algorithmic pricing should stimulate competition but, in some instances, the opposite trend has been observed: very reactive pricing strategies can discourage competition and, ultimately, lead to increased prices~\citep{Brown2022}. In addition, algorithmic pricing tends to increase price variability and unpredictability~\citep{Bertini2021}, and there are concerns as to whether it promotes collusion~\citep{CMA2018, Assad2020}.

From a modeling perspective, a major challenge in algorithmic pricing stems from the presence of multiple decision-makers with conflicting interests \textcolor{black}{ at least at two levels: \textit{producers} and \textit{customers}.} {\color{black} Here, however, keep in mind there could be additional levels depending on the depth of the incumbent supply chain.}
This complexity requires a comprehensive analysis encompassing the interactions between customers and companies and the competition among companies for market share. Historically, such competitive relationships
have been conceptualized within the framework of game theory, as evidenced by works such as \cite{Rao1972, Mesak1979}, \cite{taleizadeh2019pricing}, \cite{Gupta2021} or \cite{maihami2023ticket}. An integral aspect of this analysis is acknowledging the role of strategic consumers, whose considerations significantly influence the effectiveness of pricing policies. For instance, \cite{Cachon2010} demonstrate that static pricing strategies might eclipse dynamic ones under certain conditions. This phenomenon is attributed to the increased risk borne by customers due to larger price variability, which might drive them towards alternative companies. The competitive interplay of company pricing strategies, especially concerning their market behavior, has been extensively explored in various studies, including a detailed examination in ~\cite{Kopalle2012}.

More recently, the focus has shifted to studying the implications of companies adopting algorithmic pricing strategies. This shift in strategy has sparked significant interest, particularly concerning its potential effects on collusion, an area where academic consensus is still evolving. \textcolor{black}{For example, \cite{Miklos-Thal2019} argue that, despite its tendency to facilitate collusion, algorithmic pricing can sometimes lead to lower prices and increased consumer surplus, as it encourages companies to reduce prices during high-demand periods. By contrast, work by \cite{Calvano2020} supports the hypothesis that this practice consistently results in supracompetitive prices.} Additionally, the consideration of consumer strategies has opened up new avenues for personalized pricing models, aimed 
 to tailor prices for individual customers \citep{Choudhary2005}, a core concept in this paper.


Central to typical game-theoretic analyses in pricing is the reliance on strong common knowledge assumptions among agents \citep{hargreaves2004game}. However, this assumption can be contentious in the practical realm of competitive business, potentially leading to inappropriate solutions. 
Other authors have critically discussed common knowledge issues in various areas of management and economics. \cite{bergemann}, for instance, examine them in the context of mechanism design, while \cite{angeletos} explore their implications for economic policy.
More general critiques of common knowledge in games, such as the common prior issue in Harsanyi's doctrine, are discussed in works by \cite{sakovics} and \cite{Antos}.
\textcolor{black}{ Although insightful, these discussions adopt a different methodological approach compared to ours, based on 
Adversarial Risk Analysis (ARA, \cite{rios2009adversarial}),
to provide an alternative personalized pricing algorithmic framework}. We refer the interested reader to \cite{banksover} for a detailed conceptual comparison of ARA with other game theoretic formalisms, where ARA advantages are showcased over other frameworks.

Our proposal 
incorporates strategic reasoning about the behavior of competitors and customers while accounting for uncertainty where necessary. The focus of our analysis will be solely on static pricing, where the primary objective is to set a price that will attract a customer at a specific point in time. \textcolor{black}{ This leaves as future work the 
dynamic aspects based on integrating our approach with the above-mentioned ML approaches to enhance their strategic aspects}.
It is important here to note that this is one of the first applications of the ARA 
framework in business competition. 
\textcolor{black}{Most previous research ARA has focused on security and cybersecurity, as reflected in numerous works, including those by \cite{roponen}, \cite{insider}, \cite{gomez2024forecasting} and \cite{albert}. Recently, however, ARA has expanded into other areas, such as adversarial machine learning \citep{gallego2024protecting}, parole board decision-making \citep{joshi2024parole}, and business applications. In the business domain, ARA has primarily been applied to auctions \emph{e.g.}\  \cite{Banks2015,banksover,joshi,indio}.
}
Previous research in 
\cite{Deng2015Application} explore pricing within a remanufacturing context, focusing on the interaction between the original equipment manufacturer and several remanufacturers. Their analysis is conducted in a sequential setting without taking consumer preferences into account. Consequently, they address a structurally simpler and more specific pricing problem than the one in this paper, which diverges significantly by incorporating this last factor employing ARA. 
Importantly, from the point of view of the ARA methodology, we present  the theoretical analysis of a novel model that involves multiple agents operating at two distinct levels (\textit{producers} and \textit{customers}), with 
each of these levels encompassing different information and decision-making dynamics. 

\textcolor{black}{ The framework is presented in Section \ref{SEC: ARA} as a generic pricing template.} It is illustrated in two areas in which pricing is of major interest:  \textit{retailing} (Section~\ref{SEC:retail}) and the \textit{pension fund market} (Section~\ref{SEC:pensions}).
The first one is used to illustrate core numerical and modeling issues, representing the key ideas needed to implement the proposed method, 
 \textcolor{black}{ as well as a comparison with standard game-theoretic approaches.
The second one complements } the initial template with additional modeling complexities. 
In both cases, an underlying theoretical model is described together with case studies used to illustrate modeling and computations. Finally, Section~\ref{SEC:discussion} provides a 
  discussion and suggests open research questions. \textcolor{black}{ The generic template is
 justified theoretically in Appendix A. All the code to reproduce the experiments presented} is available at \url{https://github.com/simonrsantana/ara_pricing}. 


\section{Problem statement and solution}\label{SEC: ARA}

\textcolor{black}{ We introduce our solution approach 
  through a template for 
 the static personalized pricing problem}. Suppose that $n$ producers, denoted $ P_1, P_2, \ldots, P_n $, set their respective prices to specific initial values $ p_1, p_2,$ $\ldots, p_n $, for a particular product aiming to attract a customer ($C$, \textit{he}). The customer compares different offers and chooses his preferred one, \textcolor{black}{ where the outcomes of the final purchase depend
 on uncertain generic features $s$ modeled through the 
  random  %
 variable $S$. Examples might include observable features at the 
 time of purchase like the delivery time, unobservable endogenous features 
 at the time of purchase like the product duration, and 
 exogenous features like economic environment variables affecting
 product usage. 
 We aim to support the first producer ($P_1$, \textit{she}) to }
optimally set her price, taking into account various sources of uncertainty, which include those related to the competitors' decisions and the customer's choice. 

\begin{figure}[htb]
\centering
\includegraphics[width=0.5\linewidth]{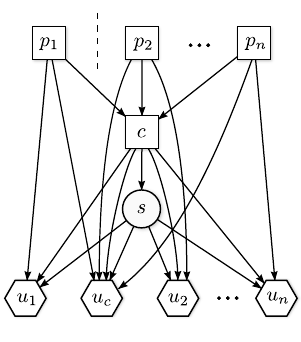}
\caption{Multi-agent influence diagram for the global pricing problem. }
\label{fig:IDglobal}
\end{figure}

Figure~\ref{fig:IDglobal} illustrates the problem as a multi-agent influence diagram (MAID) \citep{Banks2015}, where square nodes represent decisions; circle nodes, uncertainties; and, finally, hexagonal nodes represent utility evaluations.
Arrows pointing to decision nodes indicate information availability
when such a decision is made, whereas arrows pointing to chance and value 
 nodes reflect (statistical) dependence. 
 %
 In this framework, the supported agent (producer 1) is presumed to be an expected utility maximizer \citep{French}, with an associated utility function denoted as $u_1$. She operates under the belief that her competitors and the customer will also aim to maximize their expected utilities, whose respective utility functions are labeled as ${u_i}$ with $i \in \{ 2, ...,n \}$ and $u_c$, and models her uncertainty regarding the other agents' beliefs and preferences via random probabilities and random utilities. Throughout the discussion, subindexes will be used to indicate the specific agent in question, while capital letters will denote random utilities or probabilities.

\begin{figure}[hbt]
\centering
\begin{subfigure}{.32\textwidth}
  \centering
   \includegraphics[width=0.8\linewidth]{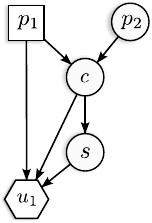}
\caption{Pricer $P_1$ problem. \vspace{0.1cm}}
\label{fig:supported}
 \end{subfigure}
\begin{subfigure}{.32\textwidth}
\centering
\includegraphics[width=0.8\linewidth]{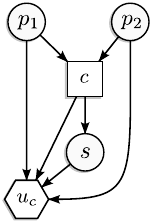}
\caption{Consumer problem. \vspace{0.1cm}}
  \label{fig:2c}
\end{subfigure}
\begin{subfigure}{.32\textwidth}
\centering
\includegraphics[width=0.8\linewidth]{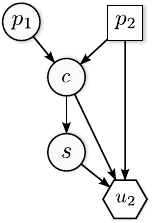}
\caption{Pricer $P_2$ problem. \vspace{0.1cm}}
  \label{fig:2b}
\end{subfigure}
\caption{The three partial problems for the pricing problem. Only two producers 
reflected.} 
\label{fig:decomposed_general_problem}
\end{figure}

  Based on the global view of the problem in Figure \ref{fig:IDglobal}, Figure \ref{fig:decomposed_general_problem} introduces three \textit{sub-problems} that represent, in order, the perspective of (\textit{a}) the supported first producer; 
    (\textit{b}) of the consumer; and, finally, (\textit{c}) of another producer\textcolor{black}{, say $P_2$}.
For the sake of simplicity, we depict these graphs only \textcolor{black} { with two producers, $P_1$ and $P_2$, although the analysis will include an arbitrary number of producers.} 
\textcolor{black}{Results used to support the correct definition of each of the three core problems framing the proposed approach are provided in Appendix \ref{App:lemmas}.}

\subsection{Supported pricer's problem}

We first analyze the decision problem faced by producer $P_1$
(Figure \ref{fig:supported}). Due to $P_1$'s lack of complete information, the 
competitor's and customer's decisions are uncertain to her and thus 
represented as chance nodes.
To solve this decision problem, we elicit the following ingredients from $P_1$. 
\begin{enumerate}
\item Her utility function, $u_1(p_1, c, s)$, modelling her preferences over possible outcomes.
\item The distribution $q_1(s \, \mid  \, c)$ modeling her uncertainty about 
\textcolor{black}{ the features affecting product outcomes given the customer's choice.}
\item The distribution $q_1(p_2\,,...\,,p_n)$ modeling her beliefs about the prices that the competitors $P_2\,,...\,,P_n $ will set for the product.
\item The distribution $q_1(c \, \mid  \, p_1, p_2\,,...\,,p_n)$ over the decision made by the customer given a set of proposed prices.
\end{enumerate}
Assessment of ingredients 1 and 2 is standard from a decision-analytic practice
 perspective; see \cite{GonzalezOrtega2018} and \cite{ohagan} respectively. Appropriate specification of the third and fourth distributions is typically more challenging as it entails strategic elements that need \textcolor{black}{ careful consideration. For improved clarity, we discuss these more in detail in Sections \ref{sec:customer_problem} and \ref{sec:competitors_problem} and in Appendix \ref{App:lemmas}.}
  For the given time, assume they are available for our analysis.

Given ingredients 1-4,  using influence diagram computations \citep{shachter} over Figure \ref{fig:supported}, producer $P_1$ should aim at finding her optimal price $p_1^*\in {\cal P}_1$ by maximising 
her expected utility
\begin{align}
\label{eq:original_problem_supported_pricer}
    & \psi_1 (p_1)  \\ & =  \sum_{c=1}^n \int \dots \int u_1(p_1, c, s) q_1(s \, \mid  \, c) 
    q_1(c \, \mid  \, p_1, ..., p_n ) q_1(p_2\,,...\,,p_n ) \, \dd s \,  \dd p_2 \dots \dd p_n  , \nonumber
\end{align}
 where  ${\cal P}_1$ represents the set of feasible prices available to her, 
and $c=i\,\,$  indicates that the consumer chooses the $i$-th product.
 In general, the optimization problem (\ref{eq:original_problem_supported_pricer}) will be solved through Monte Carlo-based decision theoretic computations as discussed in \emph{e.g.} \cite{shao1989monte, French}, \citeauthor{EKIN2022} (\citeyear{EKIN2022}, suppl. materials)  or \cite{powell}
and  
Sections \ref{SEC:retail} and \ref{SEC:pensions} illustrate.
\subsection{Customer's problem}\label{sec:customer_problem}

Let us analyze now the customer's perspective to assess the 
\textcolor{black}{ first missing strategic element,
$q_1(c \, \mid  \, p_1, p_2,..,p_n)$}. His multiple comparison problem is reflected in the setup \textcolor{black}{ in Figure \ref{fig:2c}, which, for improved readability and simplicity in the discussion,  
just reflects two producers}.

In the fundamental version of the model, the customer's decision-making process is straightforward: he opts for the product from the first producer ($ c = 1 $) if, and only if, the expected utility he derives from it surpasses \textcolor{black}{ those of the competitors' products. However,  observe that this expected utility calculation is not just about price comparison, since it also reflects other product outcomes through $s$.}  Formally, \textcolor{black}{the consumer chooses the first product ($ c = 1 $) for a given set of prices $p_1$, $p_2$, ..., $p_n$ if we have that}
\begin{eqnarray}
    \label{eq:customer_problem_pricing}
    h (p_1, p_2\,,...\,,p_n) = \int u_c(p_1, s) p_c(s \, \mid  \, c = 1) \, \dd s \,\, - 
    \max_{i=2\,,...\,,n} \left( \int u_c(p_i , s)p_c(s \, \mid  \, c = i) \, \dd s \,\right) \geq 0.
\end{eqnarray}
  \textcolor{black}{ As argued in \cite{KEENEY},} typically we do not fully know the \textcolor{black}{ consumer's utility   
  $u_c$ and probabilities $p_c$}. To overcome this, we use a Bayesian
  approach by modeling the corresponding uncertain elements as random utilities $U_c$ and random probabilities $P_c$ \citep{Banks2015}, 
  from which we obtain the (random) difference in expected utility 
\begin{eqnarray}\label{sidecars}
    H(p_1, p_2\,,...\,,p_n) = \int U_c(p_1, s) P_c(s \, \mid  \, c = 1) \, \dd s \, -
    \max_{i=2\,,...\,,n} \left( \int U_c(p_i, s)P_c(s \, \mid  \, c = i) \, \dd s \, \right). 
\end{eqnarray}
From this, we assess the probability that the customer selects the first product over the competitors' as 
\begin{equation}\label{amish}
    q_1(c=1 \, \mid  \, p_1, p_2\,,...\,,p_n) \, 
    = \, Pr( H (p_1, p_2\,,...\,,p_n ) \geq 0).
\end{equation}
 In general, $H (p_1, p_2\,,...\,,p_n)$ will have to be approximated via Monte Carlo,
by sampling from the random utility $U_c$ and probabilities $P_c$ 
and solving the corresponding multiple comparison problem. This provides
a sample from $H (p_1 ,p_2\,,...\,,p_n )$ from which we build the 
required distribution $q_1 (c \mid  p_1 ,p_2\,,...\,,p_n)$
 \textcolor{black}{ through empirical frequencies}, as the case studies will 
 illustrate.

   From a modeling perspective, $U_c$ could adopt some
 parametric form $u_c$ and the uncertainty modeled over the 
 parameters would induce the random utility. $P_c (s\mid c=1)$ could be 
 based on $p_1 (s\mid c=1)$ with some uncertainty around it, and similarly 
 for $P_c (s\mid c=i)$, $i > 1$, \textcolor{black}{ as showcased in Section \ref{SEC:pensions}}.
 Alternatively, we could base the assessment 
 of $H$ on stochastic versions of discrete choice models \citep{train}, \textcolor{black}{  with 
 distributions over their parameters} 
 as per Section \ref{SEC:retail}.


\subsection{Competitors' problems}\label{sec:competitors_problem}

   To assess the distribution $q_1(p_2\,,...\,,p_n)$ over the competitors' prices, let us  analyze the scenario from the perspective of the other producers, represented by 
$P_2$ in Figure \ref{fig:2b} in a simplified setup with just two producers. 

$P_2$'s decision-making problem is symmetrical to that of $P_1$. Thus, 
the optimal price for $P_2$ would result from maximizing her expected utility,
\begin{align}
	\label{eq:supported_pricer_problem_original}
    & \psi_2 (p_2)  \\
    = & \sum_{c=1}^n \int \dots \int u_2(p_2, c, s) q_2(s \, \mid  \, c) q_2(c \, \mid  \, p_1, p_2\,,...\,,p_n) q_2(p_1,p_3\,,...\,,p_n) \, \dd s \, \dd p_1 \dd p_3 ... \dd p_n ,  \nonumber
\end{align}
where all the functions involved have a similar interpretation as 
in \eqref{eq:original_problem_supported_pricer} but from the perspective of agent $P_2$ (and similarly for the other producers whenever the problem includes more than $2$ competitors).

In this case, the optimal price $p_2^* = \text{arg max}_{p_2} \, \psi_2 (p_2)$ is unknown to us, as we do not typically have full access to the utilities ($u_2$) and distributions ($q_2$) of the competitor $P_2$. As before, to address this issue, we take a Bayesian approach and model the corresponding unknown elements in \eqref{eq:supported_pricer_problem_original} as random utilities $U_2$ and random probabilities $Q_2$, thereby encoding our beliefs about them. This enables us to encode our beliefs regarding these elements, accounting for both pre-existing information and our \textcolor{black}{ uncertainty 
 about  } such information, both of which can be informed by the supported pricer. We then compute the \textit{random expected utility} of competitor $P_2$ through
\begin{align}
  & \Psi_2 (p_2) \nonumber \\
  = & \sum_{c=1}^n \int \dots \int U_2(p_2, c, s) Q_2(s \, \mid  \, c) Q_2(c \, \mid  \, p_1, p_2\,,...\,,p_n) Q_2(p_1,p_3\,,...\,,p_n) \, \dd s \, \dd p_1 \dd p_3 ... \dd p_n . \nonumber
\end{align}
which induces the \textit{random optimal price} for the second producer
\begin{equation*}
    P_2^* = \argmax _{p_2} \Psi_2 (p_2),
\end{equation*}
and set the desired distribution through
\begin{equation*}
    Q_1(p_2) = Pr(P_2^* \leq  p_2),
\end{equation*}
where $Q_1$ designates the \textcolor{black}{cumulative distribution function} of $q_1 (p_2 )$.

 In practice, we would obtain Monte Carlo samples from $P_2^*$ by drawing samples from the random utilities $U_2$ and probabilities $Q_2$ and finding the corresponding optimal $p_2^*$. We then use its empirical distribution function as an approximation to $ Q_1(p_2)$. This is illustrated in the case studies presented
in Sections 3 and 4.
From a modeling perspective, if we assume that the competitor's beliefs about the customer's behavior are similar to those of the supported pricer, the assessment of the random distributions $Q_i (s\vert c)$ and $Q_i(c\vert p_1, ..., p_n )$ can be based on our own distributions $q_1(s\vert c)$ and $q_1(c\vert p_1\,,...\,,p_n)$, with some additional uncertainty around them, as later exemplified.
As before, $U_i$ could adopt some
 parametric form $u_i$ and the uncertainty modeled over the 
 parameters would induce the random utility.
 To model $Q_i (p_1, p_2\,,...\,,p_{i-1},p_{i+1}\,,...\,,p_n)$, we must consider 
the $i$-th producer's beliefs about the remaining producers' prices; if $P_i$ is assumed to believe that producers set prices independently, $Q_i$ can be represented as $Q_i (p_1, p_2\,,...\,,p_{i-1},p_{i+1}\,,...\,,p_n) = \prod_{j=1, j\neq i}^n Q_i(p_j)$. Here, each element $Q_i(p_j)$ might be modeled by a parametric distribution, like a gamma distribution, where the mean represents an estimate of $p_j$ based on prior information available to all producers, and the variance regulates our uncertainty in $P_i$'s making such price estimation.
In any case, the specific formulations of random utilities and probabilities will depend on the particularities of each problem. We demonstrate some examples in the case studies below.  

In a problem involving $n>2$ producers, see Figure 1, we would use an analogous model for each competitor $P_i, i \in \{3\,,...\,,n\}$, to obtain the respective marginal distributions $q_1 (p_i)$ over their prices. Then, an assumption of independence among the producers, which is reasonable in absence of collusion, would allow us to set up the joint distribution of their prices.

\subsection{Comments}
Our basic initial hypothesis entails that $P_1$ is
a level-2 agent in \cite{stahl1994experimental} sense. Extensions to higher-level agents and cases in which the other producers and/or the customer do not maximize expected utility follow a path 
similar to \cite{oppon} albeit in a different context.
The following sections showcase the usage of this general template in two pricing problems of major interest, namely in the \textit{retailing} and \textit{pension fund market} sectors.

\section{Pricing in retailing}\label{SEC:retail}

Retailing is a major sector in modern economy. As an example, in Spain, it 
contributes to more than 5\% of the GDP \citep{ICI23} whereas it covers around 17\% of employment in the European Union \citep{empEU22}. Pricing is a critical problem, for instance, at the lower end of the fashion sector, with intense competition among international brands, where  
  price is a major driver of consumer behavior. 
 
In this section, we assume that the customer bases his decision solely 
   on price, excluding other variables such as product performance, retailer marketing efforts, product presentation and exposition, brand recognition, status, or past interactions between the consumer and the producer. This simplification is made since items are considered highly similar across producers. The model employed here is thus a streamlined version of the general template from Section \ref{SEC: ARA} \textcolor{black}{ not including an $S$ node}.  This particular case allows us to delve into the specifics of modeling, \textcolor{black}{ numerical, and algorithmic details within a relevant domain application,
while facilitating comparisons concerning knowledge assumptions}.
 
\subsection{Problem formulation} 
\label{sec:model_formulation}

Consider supporting retailer $P_1$ against several competitors. A typical context would be a fashion retailer who forecasts a certain amount of sales in a period and, not meeting such forecast, decides to change its price as a way to attract customers. 
For the most part, we discuss a single-competitor case, although the formulation is easily extended to include more competitors, as Sections 2.2 and
2.3 
 discussed.

\begin{figure}[htb]
\centering
\includegraphics[width=0.35\linewidth]{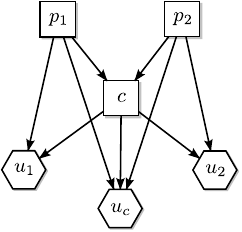}
\caption{Global pricing problem in retailing with two producers.}\label{fig:global_pricing_problem}
\end{figure}

Figure \ref{fig:global_pricing_problem} presents the problem from a global perspective. The major difference with Figure \ref{fig:IDglobal} is the absence of node $S$, as products are considered essentially homogeneous. 
From Figure \ref{fig:global_pricing_problem}, we would deduce three partial figures \textcolor{black}{ as we did from  Figure \ref{fig:IDglobal}.} Due to similarity, we omit them and just comment upon their handling, when supporting retailer $P_1$ in deciding its optimal price.

\noindent Similar to problem (\ref{eq:original_problem_supported_pricer}), for the first retailer,  price $p_1$ should be set at  
\begin{equation}\label{eq:retailer_1_substituted_problem}
    p_1^* = \argmax_{p_1} \psi_1(p_1) =  \argmax_{p_1} \sum_{c=1}^2 \int u_1(p_1, c) q_1 (c \, \mid  \, p_1, p_2) q_1(p_2) \, \dd p_2. 
\end{equation}
Symmetrically, akin to \eqref{eq:supported_pricer_problem_original}, retailer $P_2$ would aim to find the value $p_2$ maximizing his expected utility.
However, as before, since $P_2$ is a competitor, we assume we do not have complete information about her preference $u_2$ and belief $q_2$ models,
and adopt random utilities $U_2$ and random probabilities $Q_2$ to forecast $P_2$'s decision through the (random) optimal price 
\begin{equation}\label{eq:retailer_decision_problem}
  P_2^* = \argmax_{p_2}  \Psi_2(p_2) = \sum_{c=1}^2 \int U_2 (p_2,c) Q_2(c \, \mid  \, p_1,p_2) Q_2(p_1) \, \dd p_1 \, ,
\end{equation}
making  
\begin{equation*}
    Q_1(p_2) = Pr(P_2^* \leq  p_2).
\end{equation*}



\noindent The consumer must then choose between the products from $P_1$ and $P_2$. Taking into account \eqref{amish}, this decision is modeled by the first retailer through the pairwise comparison problem
\begin{equation*}
    q_1(c=1 \, \mid  \, p_1, p_2) = Pr(U_c (p_1) \geq U_c (p_2)),
\end{equation*}
where $U_c$ is a random utility function modeling the partially known preferences
 of the consumer. \textcolor{black}{ Finally, we will also have that 
 \[ 
 q_1(c=2 \, \mid  \, p_1, p_2) = 1 - q_1(c=1 \, \mid  \, p_1, p_2) ),
 \]
since now the consumer must choose between the only two available options.}
%
\subsection{Modelling and algorithmic details}

This section illustrates key modeling and algorithmic steps in the problem presented. A typical scenario assumes that producer $P_1$ has a forecast on product sales referring, for example, to the next week. This is represented by, \emph{e.g.},   a $0.9$ predictive interval $[f_1, f_2]$  \citep{westharrison}. 
Suppose that actual sales $x$ are such that $x < f_1$ and, therefore, $P_1$ decides to intervene over prices as a means to attract customers and increase sales. Assume her current price is $\hat{p}_1$ and $v_1$ is her internal product valuation, which, in principle, constrains her price to be $p_1 \geq v_1$; \textcolor{black}{ as an example, $v_1$ can subsume } the production, transportation, marketing, and distribution costs for the product at hand. Her competitor's current price is $\hat{p}_2$ with valuation $v_2$, this one being only partially known. As mentioned, customers are assumed to make the purchasing decision solely attending to its price.
We solve this problem as described in Section \ref{sec:model_formulation}, introducing core modeling elements.

Assuming the sale of a non-perishable product that is not expensive,
 \textcolor{black}{ producer $P_1$'s utility function can reflect a \textit{risk-neutral} behavior } 
  adopting the form
\begin{equation}
\label{eq:utility_setup_supported_pricer}
    u_1(p_1, c) = 
    \begin{cases}
    p_1 - v_1 \quad & \text{if } c = 1,\\
    0 \quad & \text{if } c = 2.         
    \end{cases}
\end{equation}
For more luxurious products, we could incorporate a risk-averse component in the utility function. For a perishable product, we would substitute $0$ by $-v_1$ to penalize the case in which $P_1$ fails to sell the product
by the required deadline.

For simplicity, for the distribution $q_1(c \, \mid  \, p_1,p_2)$ over the customer decision depending on the two prices offered, instead of the simulation-optimization approach presented in \ref{sec:model_formulation}, \textcolor{black}{ we use a probit discrete choice model 
  from the 
consumer behavior economics literature, see e.g.\ 
 \cite{train}},
    \begin{equation}\label{eq:customer_decision_model_including_uncertainty_R1}
        Pr(c=1 \, \mid  \, p_1, p_2, \sigma_1) = 1 - \phi\left(\frac{p_1 - p_2}{\sigma_1 }\right),
    \end{equation}
    where $\phi$ is the standard normal cumulative distribution function
   \textcolor{black}{  and $\sigma_1$ may be seen as a description of how firm
    are the preferences of the consumer.}
\textcolor{black}{ Observe that in (\ref{eq:customer_decision_model_including_uncertainty_R1}),
    if $p_1 < p_2$, then $Pr(c=1 \, \mid  \, p_1, p_2, \sigma _1 ) > 0.5\,\,$ 
         reflecting the fact that lower price makes
         more likely product purchase.}    
     We assume uncertainty about its standard deviation $\sigma_1$, incorporating to (\ref{eq:customer_decision_model_including_uncertainty_R1}) an inverse-gamma  prior
        \begin{eqnarray}\label{kakawoo}
               \sigma_1^2 \sim \Gamma^{-1}(\alpha_1, \alpha_2),
        \end{eqnarray}
        with parameters $\alpha_1$ and $\alpha_2$ adapted to  
        the market segment to which the customer belongs. Under such prior, it is easy to prove that 
        \[ 
        Pr(c=1 \, \mid  \, p_1, p_2) = 1 - Pr\left(T \leq \sqrt{\frac{\alpha_1}{\alpha_2}} \cdot (p_1 - p_2)\right) ,\]
        where $T$ follows a $t$-distribution with $2 \alpha_1$ degrees of freedom.
          

    To deal with the strategic component when forecasting $q_1(p_2)$, we model it considering problem \eqref{eq:retailer_decision_problem}. The utility function of $P_2$ adopts the form 
    \begin{equation*}
        u_2(p_2, c) = \begin{cases}
        0 \quad &\text{if } \: c = 1, \\
        p_2 - v_2 \quad &\text{if } \: c = 2.
        \end{cases}
    \end{equation*}
    We could add stochastic elements by including some uncertainty about $v_2$
    (say a uniform distribution around $\hat{v_1}$) 
    reflecting  lack of knowledge about $P_2$'s  production processes,
        and an eventual risk aversion coefficient (\emph{e.g.},  should the product be expensive).
    
    For $q_2(p_1)$, \emph{i.e.} the model of what $P_1$ believes $P_2$ thinks about which will be $P_1$'s price, we consider the general density 
        \begin{equation}
    \label{eq:q_2_p_1_setup}
        q_2(p_1) = \begin{cases}
        0, &\quad \text{if } \: p_1 < v_1; \\
        \frac{n+1}{(\hat{p}_1 - v_1)^{n+1}} \cdot (p_1 - v_1)^n, &\quad \text{if } \: v_1 < p_1 < \hat{p}_1; \\
        0, &\quad \text{if } \: \hat{p}_1 < p_1 . 
        \end{cases}
    \end{equation}
    This choice enables us, on the one hand, to reflect basic business information concerning a feasible price range  $[v_1, \hat{p}_1] $, \textcolor{black}{ so as to not deviate much from the current price $\hat{p}_1$,} and, 
    \textcolor{black}{ on the other, facilitates simulating from this 
    distribution via the inverse transform sampling method}. Some additional uncertainty 
    could be modeled through the parameter $n$. 
    
	Finally, for $Q_2(c \, \mid  \, p_1, p_2)$ we use a symmetric setup to that of $P_1$, that is
    \begin{equation}
    \begin{aligned}
    \label{eq:Q_2_for_c_given_p1_p2}
        Pr(c=1 \, \mid  \, p_1, p_2) & = 1 - \phi\left( \frac{p_1 - p_2}{\sigma_2} \right), \\ 
        \sigma_2 ^2  & \sim \Gamma ^{-1} (\beta_1, \beta_2),
    \end{aligned}
    \end{equation}
    with the distribution of $\sigma_2$ typically reflecting bigger uncertainty than that of $\sigma_1$ in (\ref{kakawoo}).
    

The required computations are then implemented as follows.
 First, the objective function of the primary optimization problem \eqref{eq:retailer_1_substituted_problem} is approximated by Monte Carlo through
\begin{align}
\label{eq:psi_p1}
  \psi_1 (p_1) &= \sum_{c=1}^2 \int u_1(p_1, c) q_1(c \, \mid  \, p_1, p_2) q_1(p_2) \dd p_2   \nonumber \\
    &= (p_1 - v_1) \int Pr(c=1 \, \mid  \, p_1, p_2) q_1(p_2) \dd p_2  \nonumber \\
    & \simeq (p_1 - v_1) \left( \frac{1}{N_1} \sum_{i=1}^{N_1} Pr(c=1 \, \mid  \, p_1, p_2^i ) \right) \nonumber \\ 
 & =  (p_1 - v_1) \left( 1 - \frac{1}{N_1} \sum_{i=1}^{N_1}
     Pr\left(T \leq \sqrt{\frac{\alpha_1}{\alpha_2}} \cdot (p_1 - p_2^i)\right)   \nonumber \right)  \\
    & \equiv \widehat{\psi}_1 (p_1),
 \end{align}
where $T$ follows a  $t$-distribution with $2 \alpha_1$ degrees of freedom
and $\{p_2^i\}_{i=1}^{N_1}$ is a sample  from $q_1(p_2)$. 
Then, we solve for
\begin{equation}
\label{eq:optimal_p1}
    \max \,\, \widehat{\psi}_1 (p_1) \quad \text{s.t.} \quad p_1 \in [v_1, \hat{p}_1]
\end{equation}
with a univariate optimization routine. 

To obtain a sample from $q_1(p_2)$, we use \eqref{eq:retailer_decision_problem} and proceed as in \eqref{eq:psi_p1}
\begin{align*}
    \Psi_2(p_2) &= \sum_{c=1}^2 \int U_2(p_2, c)Q_2(c \, \mid  \, p_1, p_2) Q_2(p_1) \dd p_1 \\
     & \simeq (p_2-v_2)
\left( 1-  \frac{1}{N_2 }  \sum_{i=1}^{N_2}  Pr\left(T \leq \sqrt{\frac{\beta_1}{\beta_2}} \cdot (p_1^i - p_2)\right)  \right)  \\
    & \equiv h(p_2), 
\end{align*}
for samples $\{p_1^i\}_{i=1}^{N_2}$ from \eqref{eq:q_2_p_1_setup}, 
where $T$ follows now a $t$-distribution with $2 \beta_1$ degrees of freedom.
Optimizing $h(p_2)$ provides us with one sample from $q_1(p_2)$. 
This process is repeated as 
 needed to attain the required precision in the Monte Carlo approximation \eqref{eq:psi_p1}.
 The approach follows the implementation in Algorithm \ref{alg:generate_p2}.

\begin{algorithm}
\caption{ \hspace{0.5cm} \textbf{\texttt{sample\_p2}} \hfill Sampling $N_1 $ times from  $q_1(p_2)$}\label{alg:generate_p2}
\begin{algorithmic}
\State \textbf{input} $N_1 $, $N_2 $, $\widehat{p}_2$, $v_2$, $\beta_1$ and $\beta_2$
\For{$h = 1$ to $N_1 $} \Comment{$N_1$ samples from $q_1(p_2)$}
    \For{$i=1$ to $N_2$}  \Comment{$N_2$ samples from $p_1$}
        \State Sample $p_1^i \sim q_2(p_1)$
    \EndFor
    \For{ $p_{2}^k \in$ grid$(v_2, \widehat{p}_2)$}
        \State Compute
        $$
        h(p_{2}^k) = \frac{1}{N_2 } \left( p_{2}^k-v_2 \right) 
        \sum_{i=1}^{N_2}    \left[1 - Pr\left(T \leq \sqrt{\frac{\beta_1}{\beta_2}} \cdot (p_1^i - p^k_2)\right) \right]
        $$
    \EndFor
    \State Set $p_2^{Sample, h} =  \argmax_{p_{2}^k} h (p_{2}^k)$
\State \textbf{return} $p_2^{Sample}$ \Comment{Return the final sample}
\EndFor
\end{algorithmic}
\end{algorithm}

Finally, we combine all the ingredients in Algorithm \ref{alg:pricing_algorithm}, which allows us to obtain the optimal pricing value $p_1^*$ for $P_1$ using grid search for optimization purposes.

\begin{algorithm}
\caption{ \hspace{0.5cm} \textbf{\texttt{optimal\_price\_p1}} \hfill Obtain optimal $p_1$}
\label{alg:pricing_algorithm}
\begin{algorithmic}
\State \textbf{input} $N_1 $, $N_2 $, $\widehat{p}_2$, $v_2$, $\alpha_1$, $\alpha_2$,  $\beta_1$ and $\beta_2$
\State \textbf{run} \texttt{sample}$\_p2(N)$ with input $N_1 $, $N_2 $, $\widehat{p}_2$, $v_2$, $\beta_1$, $\beta_2$ \Comment{$N_1$  $q_1(p_2)$ samples}
\For{$p_1^j \, \in \, \text{grid}[v_1, \widehat{p}_1]$}
    \State Compute 
    $$
    \widehat{\psi}(p_1^j) = \frac{1}{N_1 }  \left( p_1^j - v_1 \right) \sum_{i=1}^{N_1} 
    \left[1 - Pr\left(T \leq \sqrt{\frac{\alpha_1}{\alpha_2}} \cdot \left(p_1^j - p_2^{\text{Sample},h  } \right)\right)  \right]
    $$
\EndFor
\State \texttt{opt\_price} = $\argmax_{p_1^j} \widehat{\psi}(p_1^j)$ 
\State \textbf{return} \texttt{opt\_price} \Comment{Return optimal price}
\end{algorithmic}
\end{algorithm}

\subsection{Case}

We analyze some practical cases in a scenario where a retailer wants to determine a new price for a given product. Table \ref{tab:init_params} presents the parameters required to set up the model. The number of samples refers to both $N_1 $ and $N_2$ in Algorithm \ref{alg:generate_p2}; $100$ samples provided 
sufficient stability for the results showcased.
For the samples from $q_2(p_1)$,  we select $v_1 = 5$, $\hat{p}_1 = 50$ and $n = 2$. The prices explored are expressed in generic units and range above and below the product cost and its initial price. Finally, in all experiments we employ $\alpha_1 = \alpha_2$ and $\beta_1 = \beta_2$ for the respective prior distributions of $\sigma_1^2$ and $\sigma_2^2$, thus imposing similar uncertainty about the customer's behavior for both the supported pricer and the competitor for a given set of prices.  This setup is common to the three cases below unless stated otherwise. 

\begin{table}[htb]
    \caption{Values for parameters in retail cases}
    \label{tab:init_params}
    \centering
        \begin{tabular}{c|c}
            \hline
            Initial price $p_1$  for $P_1$ & $40$ \\ 
            Initial price $p_2$  for $P_2$  & $40$ \\
            Product cost for $P_1$ & $5$ \\
            Product cost for $P_2$ & $5$ \\
            Number of samples & $100$ \\ 
            Price range explored for $P_1$ & $[5, 50]$ \\
            \hline
        \end{tabular}
\end{table}

\begin{figure}[htb]
\centering
\includegraphics[scale=0.6]{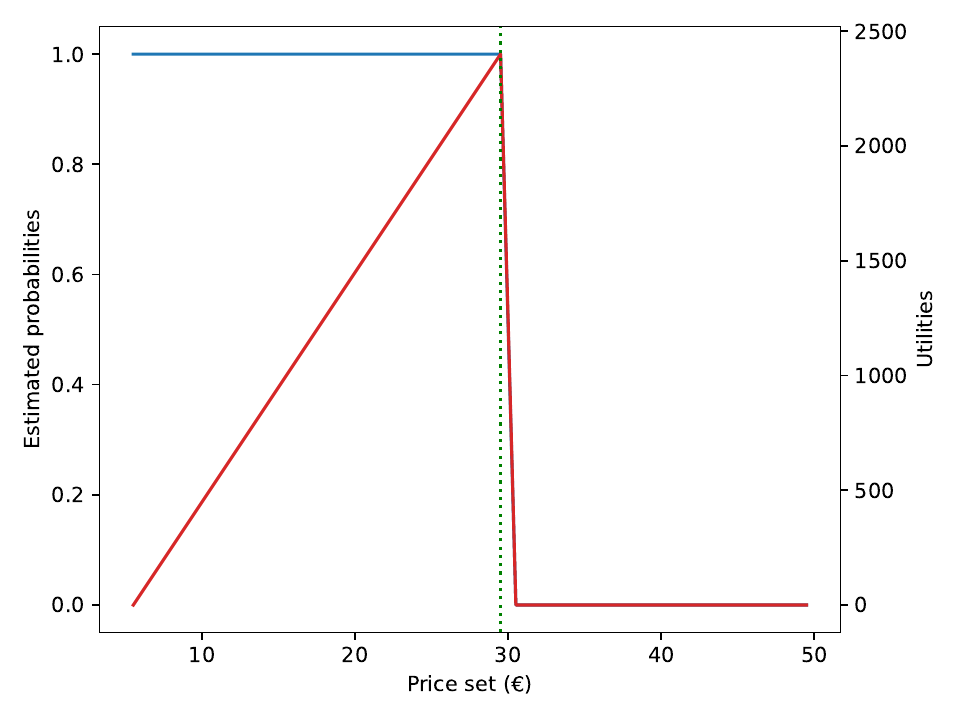}
\caption{Results for first case; 
 $p_2 = 30$, $\sigma = 0.01$. 
 Price selected, $29.5$ (best viewed in color).}
\label{fig:ejemplo_caso_1}
\end{figure}

\begin{figure}[htb]
\centering
\includegraphics[scale=0.6]{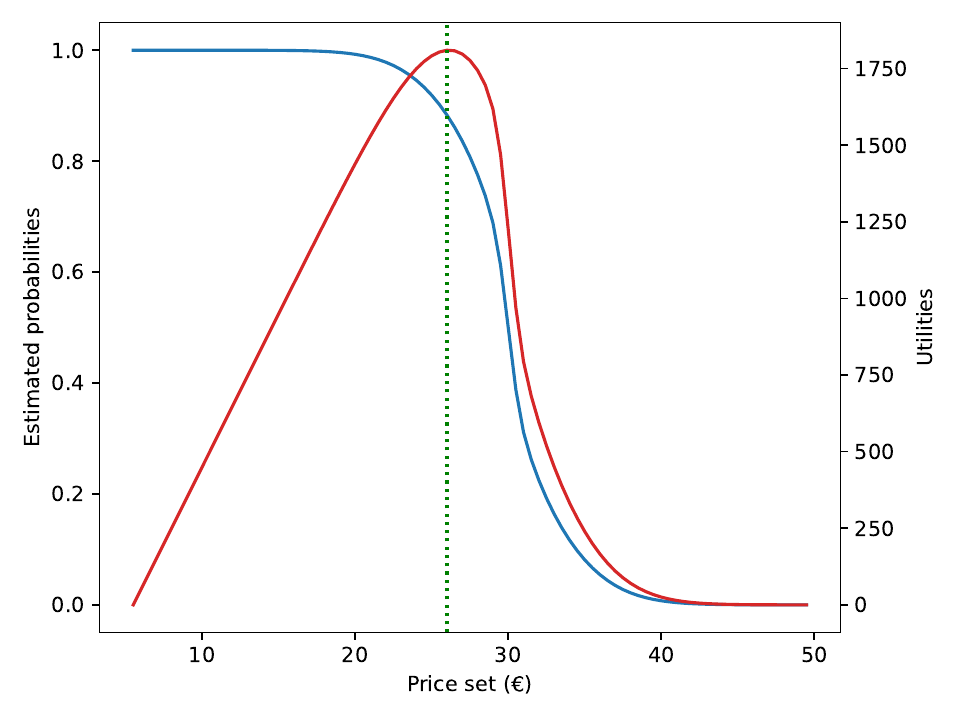}
\caption{Results for second case. 
 $p_2 = 30$; $\sigma^2_{1,2}$ sampled from inverse-gamma priors. Optimal price, $26$ (best viewed in color).}
\label{fig:ejemplo_caso_2}
\end{figure}

\begin{figure}[htb]%
\centering
\includegraphics[scale=0.6]{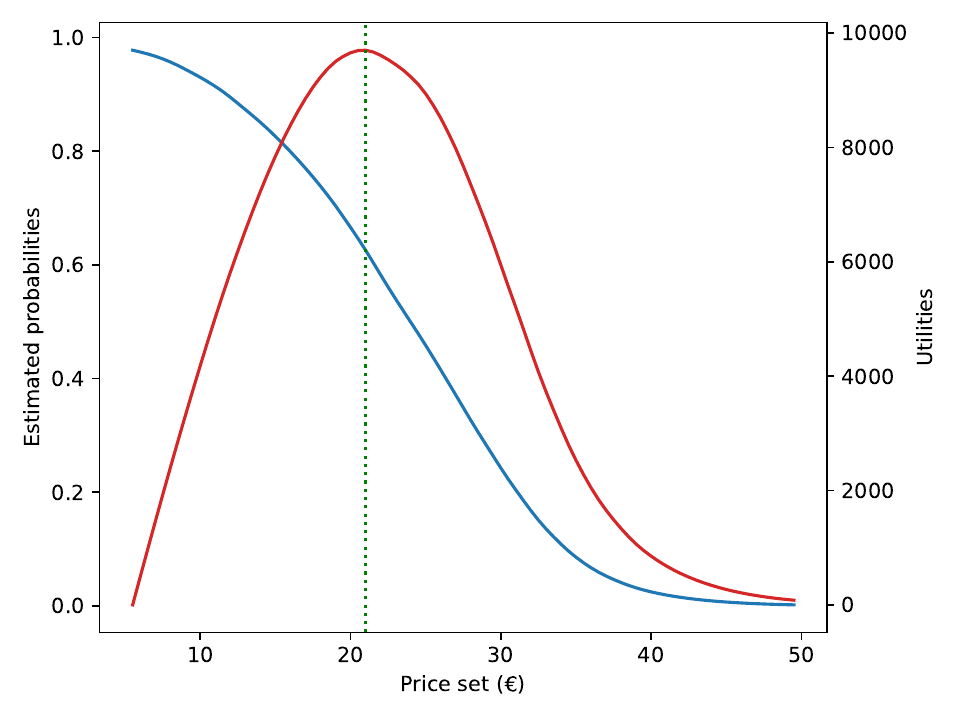}
\caption{Results for the third case. Price selected is $21$ (best viewed in color)}  
\label{fig:ejemplo_caso_3}
\end{figure}

 Figures \ref{fig:ejemplo_caso_1}, \ref{fig:ejemplo_caso_2} and \ref{fig:ejemplo_caso_3} summarize the results for three versions of the problem; the red line illustrates the expected utilities of the supported retailer, the blue line represents the estimated probability of the customer making a purchase based on the price -- considering potential competitor offers, and the vertical green dotted line signifies the suggested optimal price, namely, the estimated maximum expected utility price. 
Each case progressively relaxes common knowledge assumptions, 
\textcolor{black}{ with the first one corresponding to the standard 
game-theoretic framework, 
whereas the third one corresponds to the proposed, more
 realistic, framework in which } the supported agent lacks access to the beliefs and preferences of other agents.

\textcolor{black}{ Importantly, observe that the three scenarios, which vary in their levels of uncertainty, result in significantly different pricing strategies, highlighting} the importance of gathering precise information about the other agents, as a \textcolor{black}{ proper characterization 
  of uncertainty} may lead to improved margins and benefits.

\paragraph{\textbf{Case 1. Benchmark. \textcolor{black}{Common knowledge 
 with known competitor's price and deterministic customer's behavior}}}
 \textcolor{black}{Consider first} a simplified version of the problem where the supported agent has complete knowledge about the competitor's beliefs and utility function. This allows her to compute $P_2$'s optimal price, which we assume to be $p_2 = 30$. Similarly, we presume a low level of uncertainty regarding the customer's decision by setting $\sigma = 0.01$, thus effectively making the customer buy the product from retailer $1$ if $p_1<p_2$, and from retailer $2$ otherwise (\emph{i.e.} deterministic behavior). \textcolor{black}{ This will make the model only need to select the best price according to \eqref{eq:original_problem_supported_pricer} and, therefore, we only require a $p_1$ slightly smaller than $p_2$.} Since we explore the $[5,50]$ range in $0.5$ increments, the price directly below $p_2 = 30$ is $29.5$, which is the one selected by the model, as Figure \ref{fig:ejemplo_caso_1} shows. Thus, the expected solution under common knowledge is selected in a principled way, maximizing the expected utility. 
\hfill $\triangle$

\paragraph{\textbf{Case 2. Known competitor's price and uncertain customer's behavior}} We extend the previous case by adding uncertainty about the customer's decision. The competitor's price remains at $p_2=30$, but we introduce some variability in the customer's behavior demanding sampling for $\sigma _1^2$ and $\sigma_2^2$. Introducing uncertainty about the customer leads to a lower optimal value for the price of the product, now at $26$ with a purchase probability of 88\%, compared to that obtained in the first case ($29.5$ with a purchase probability of nearly 100\%). \hfill$\triangle$

\paragraph{\textbf{Case 3. Uncertain competitor's and customer's behavior}} We now implement the proposed model without common knowledge assumptions. We take $\alpha_1=2$ and $\beta_1 = 0.5$. We obtain each $p_2$ optimizing \eqref{eq:retailer_decision_problem} using the samples from $q_2(p_1)$. 
Then, we estimate the probability that the customer buys the product for each possible $p_1$ and choose the optimal price as the solution of \eqref{eq:optimal_p1}. Figure \ref{fig:ejemplo_caso_3} summarizes 
the results. The optimal price in this case ($21$) is significantly smaller than the previous two, as is the estimated probability of the customer buying the item at this price ($63\%$).  \hfill $\triangle$

\vspace{0.1in}
 
\noindent Observe that the three cases display a common pattern concerning the estimated probabilities and expected utilities for the extremes of the price range explored, 
serving as a sanity check for the model. Lower prices for the product increase the probability that the customer will purchase it. Beyond a certain threshold, the supported pricer's expected utility decreases due to lower earnings. Conversely, higher prices reduce customer probability of purchase. Eventually, despite a greater margin between the selling price and production costs, the expected utility diminishes to zero. \textcolor{black}{ Thus, importantly, the proposed model 
  is highly interpretable, therefore potentially providing relevant support} in practical scenarios.


\section{Complex pricing in the pension fund market}\label{SEC:pensions}
Consider now a more complex instance of Section \ref{SEC: ARA} template through a pricing problem in the pension fund market. In many economies, pension funds constitute an important complement to state-sponsored pensions, with fierce competition among the involved agents. As an example, in Spain, the estimated value of this market is 11.000M\euro{}, with nearly 10M active pensions in 2023 \citep{pensionesESP23}. The scenario we consider is that of a bank branch director who, based on a
benchmark product, is given some flexibility around such product features to attract a potential customer of interest.

The increased complexity of the problem is caused by 
the larger number of variables and parameters that need to be chosen and decisions to be made, \textcolor{black}{including: the fixed return offered, minimum permanence time, and penalty for
non-compliance which conform the pricing decision in this 
 case}, 
 the \textcolor{black}{impact of time on uncertainty regarding the customers' permanence in the fund for the initially contracted years}, and, finally, the presence of covariates characterizing the 
customer's evolution, which further \textcolor{black}{ impact} personalization. {\color{black} All these are critical aspects of the problem. The proposed model enhances interpretability and transparency by requiring an explicit definition of hypotheses around these potentially complex points, potentially facilitating the practical adoption of these techniques \citep{bibal2021legal}.}

\subsection{Problem formulation}

\begin{figure}[htb]
\centering
\includegraphics[width=0.6\linewidth]{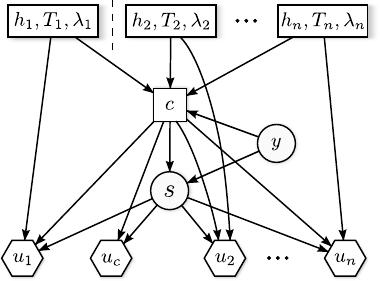}
\caption{Pension fund problem}\label{fig:complete_model_fund_market}
\end{figure}

The MAID in Figure \ref{fig:complete_model_fund_market} extends that of Figures \ref{fig:IDglobal} and \ref{fig:global_pricing_problem} by capturing the additional complexity in this context. Products are not only characterized by the entailed \textcolor{black}{ rate} ($h$), but also by the minimal time ($T$) a customer has to stay in the fund without facing a penalty ($\lambda$) for early withdrawal.  The $i$-th bank
 determines its values,
denoted $(h_i, T_i, \lambda_i)$, which collectively constitute their pricing decision for the pension product.
 As mentioned, there is a need to take into account various customer characteristics that can influence the final offer.  We refer to these variables as ``node $y$" in Figure
\ref{fig:complete_model_fund_market}. These attributes could include factors such as socio-economic status or the amount of capital the customer plans to invest in the pension fund. By including these variables, we gain deeper understanding of how different customers might react to our proposal,
\textcolor{black}{  beyond the information modeled through $u_c$}. Moreover, such variables can assist us in predicting how long a customer might stay with the pension fund 
should they accept an offer, \textcolor{black}{ represented by $s$, 
which is the 
core uncertain factor affecting the pension outcome for 
the provider and requires taking into account dynamic aspects in their 
 handling as we shall see. }


\subsection{Modeling details}

Suppose a bank branch manager (\textit{she}) has the capacity to offer a fixed return fund product with yearly return $h_1\in [h_{1,l}, h_{1,u} ]$, with this range defined by the central organization. Typically, $h_{1,l}$ will be the return advertised in the organization's marketing, whereas $h_{1,u}$ will be hidden from the customer. {\color{black} This maximum return that the bank is willing to offer will typically be based on factors such as their funding needs, growth targets, economic conditions, consumer-specific information, and 
rates offered by the competition.} The product is also characterized by $T_1$, the minimum number of years the customer should maintain the selected pension fund to avoid a penalty $\lambda_1$, imposed when 
the required permanence is not respected. A potential customer with a capital $x$ and socio-demographic features $y$, considers applying for the product. We aim to support the manager in deciding what final offer $h_1$ to make to convince such customer. This is straightforwardly extended to the case of optimizing the offer in terms of $T_1$ and $\lambda_1$ alongside $h_1$.

{\color{black}
Let $ p(h_1 \mid y) $ be the probability that a customer with covariates $ y $ will accept the offer $ h_1 $. If $ z $ are the expected yearly earnings (as a rate over the capital) of the branch, then the bank's benefit for the next year would be $ (z - h_1) \cdot x $ if the customer accepts, which happens with probability $ p(h_1 \mid y) $. Otherwise, these will be $ 0 $ if the customer does not accept, which occurs with probability $ 1 - p(h_1 \mid y)$. Notice that we assume the offered pension has no commission for the client. Therefore, the bank's earnings are solely determined by the difference between the expected yearly yield on the client's capital ($z \cdot x$) and the amount paid back to the customer ($z \cdot h_1$).
}

 
We assume the branch manager is interested in maximizing the expected utility for the next year associated with said customer {\color{black} (longer-term perspective would consider the problem for several years)}. Without loss of generality, we consider that $u_1 (0)=0$, \emph{i.e.}\ the utility obtained if the customer does not accept the offer is null. Thus, the manager should solve for
\begin{equation} \label{eq:pension_branch_manager_original_problem}
h_1^* = \argmax_{h_1\in [h_{1,l}, h_{1,u}]} \, u_1 ((z-h_1) \cdot x) \cdot p(h_1 \vert y) ,
\end{equation}
where $u_1$ designates the utility function of the branch manager.
Solving (\ref{eq:pension_branch_manager_original_problem}) requires assessing:
  \textcolor{black}{ the utility $u_1$, a standard modeling practice
   in decision analysis \citep{French};  
    the expected earning $z$, a standard problem in finance
    \citep{lamont}; and, the somewhat less standard acceptance probability $p(h_1 \vert y)$, given its strategic aspects, which we address next.}

 To estimate this last term, in line with Section \ref{SEC: ARA}, we model the customer's problem, reflecting three possible scenarios, where, for simplicity, we consider that the customer decides to select either our organization or one of the competitors for a pension fund:
\begin{enumerate}
    \item He accepts and stays the required $T_1$ years, earning $ (1+h_1)^{T_1} x$.
    \item He accepts, but stays only $t_1 (<T_1) $ years. 
    He then earns $ (1+h_1)^{t_1}  x - \lambda_1$.
    Letting $q(i \mid  y)$ be the probability that the customer stays until the $i$-th year, the probability that he will stay $T_1$ years in the fund is $q (T_1 \mid  y) = 1 - \sum_{j=0}^{T_1 - 1} q(j \mid  y)$.
    \item He does not accept our offer and adopts the $i$-th competitor's product, characterized through parameters $h_i$, $T_i$ and $\lambda _i$.
\end{enumerate}
As before, denoting by $u_{c}$ the customer's utility function, the expected utility that he would receive if he adopted our product is  
\begin{eqnarray}
\label{eq:client_utility_our_product}
\psi_1 (h_1 \mid  y) & = & \left( 1 - \sum_{j=1}^{T_1 - 1} q(j \mid  y) \right) u_{c} \left( 
(1+h_1) ^{T_1}  x\right)  \nonumber \\ 
&&  + \sum _{j=1} ^{T_1 - 1} q(j \mid  y) \, u_{c} \left( (1+h_1) ^j  x-\lambda_1  \right) + g (T_1),
\end{eqnarray}
where $g (T_1)$ represents the evaluation for having the capital available at time $T_1$, typically a decreasing function on $T_1$ (\emph{i.e.},  the customer assigns less utility to potential benefits the further they are in the future). This is to be compared with the expected utility that he would get by adopting the $i$-th competitor's product with an 
expression similar to \eqref{eq:client_utility_our_product}, with subindex $i$ replacing subindex $1$.

  Should we know $u_{c}$ and $g$, we would find the customer's optimal decision as the product maximizing $\psi_i (h_i \mid  y) $. Since this is not the case, we use random functions $U_c$ and $G$ which would give us the optimal random decision and, consequently, the required probability through
\begin{equation}\label{monday}
 p(h_1 \mid  y) = Pr _{U_c, G} \left(\Psi_1 (h_1 \mid  y) \geq \max _{i\geq 2} \Psi_i (h_i \mid  y) \right),
 \end{equation}
 where, 
 typically, we would need (\ref{monday}) for all values $h_1$. In this case, we can compute all those probabilities in a grid and interpolate. 

 Finally, note that if  it is reasonable to assume that all competitors behave identically and independently, we can write
\begin{equation} \label{eq:n_equal_competitors}
 p(h_1 \mid  y) = \left( Pr _{U_c, G} \Big( \Psi_1 (h_1 \mid  y)  \geq \Psi_2 (h_i \mid  y)  \Big) \right)^{n-1}.
\end{equation}
  Observe that, in this case, for any $h_1$ such that
$Pr _{U_c, G} (\Psi_1 (h_1 \mid y)  \geq \Psi_2 (h_i \mid y)  ) < 1$, we have that $p(h_1 \mid y)$ will tend to 0 as the number of competitors increases.

\subsection{Case}
Let us illustrate these ideas with a case study. 
\textcolor{black}{ The numerical parameters reproduced typical Spanish market figures. The first example with just one competitor and no covariates will serve as a benchmark for the other 
examples}. 

\paragraph{\textbf{Case 1. Benchmark}}\label{kkuro}

Suppose that:
\begin{itemize}
	\item The potential customer's capital is $x = 30$K \euro{}. 
	\item Analysts estimate the bank's expected earning rate at $7\%$ ($z=0.07$).
	\item Marketing specialists suggest a nominal value of $h_{1,l} = 0.025$ (i.e., 2.5\%) for the return $h_1$.
 To cover all possibilities, we set the upper limit at $h_{1,u} = 0.07$ (return 7\%, thus predicting zero net gains).
 	\item The entity imposes that the potential customer stays $8$ years to avoid being penalized, with a penalty of $80\%$ of the bonus accumulated up to that point (in \euro{}).
 	\item Based on previous data on other customers, the analysts estimate the probability that the customer leaves in years $\{1, 2, 3, 4, 5, 6, 7\}$ respectively by $\{0.15, 0.05, 0.04, 0.03, 0.02, 0.01, 0\}$ (the remaining probability represents the probability of the customer completing the $8$-year period in the pension fund).
  \end{itemize}
On the other hand, regarding the parameters of the competitor, assume that:
\begin{itemize}
    \item With no loss of generality, we set the same permanence period, penalty, and exit probabilities for the customer for each year. These values can be changed to reflect other scenarios. 
    \item \textcolor{black}{ The nominal value for $h_2$ is sampled from the set 
    $\{0.025, 0.03, 0.035, 0.04, 0.045, 0.05,$ $0.055, 0.06, 0.065, 0.07  \}$ 
    with probabilities $\{ 0.05, 0.1, 0.2, 0.2, $ $ 0.15, 0.1, 0.1, 0.05, 0.05\}$, respectively.}
\end{itemize}
\noindent
Finally, assume constant absolute risk aversion (CARA) utility 
functions  \citep{GonzalezOrtega2018}, defined as 
$u(x) = 1 - \exp (-\rho x )$, where $\rho$ denotes the risk-aversion parameter. 
 We do not make use of a $G$ function in this case. 
The uncertainty about the risk-aversion coefficient of the customer 
is modeled with a uniform prior on 
an interval $[\rho_c^1, \rho_c^2]$, which will depend on the entity, since one customer may perceive some entities as lower or higher risk options compared to others. In this particular example, we use $\rho_c^1 = 0.85$ and $\rho_c^2 = 0.95$ to model a potential customer with some 
risk-aversion behavior shared across all entities. 

\begin{figure}[htb]
\centering
\includegraphics[width=0.6\linewidth]{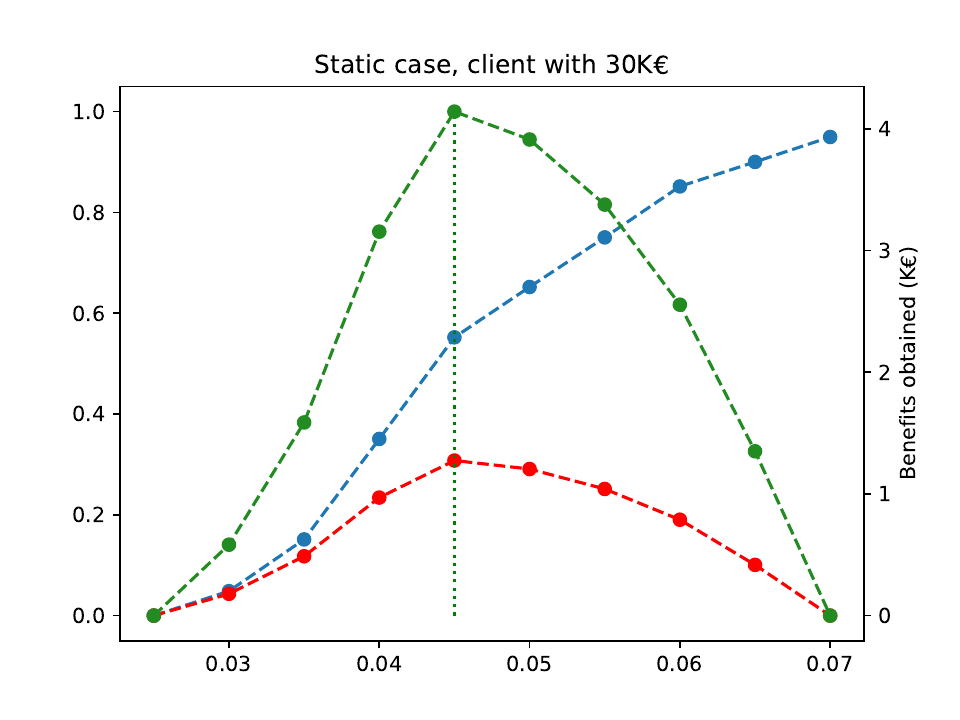}
\caption{Results for $h_1 \in [0.025, 0.07]$. Left y-axis represents the estimated acceptance probability and standardized expected utility (best viewed in color). 
}\label{fig:static_plot}
\end{figure}

\textcolor{black}{ Figure \ref{fig:static_plot} depicts the estimated probability that the customer accepts the offer (blue), the expected utility (red), and the expected benefits for the bank for each offer 
(green) for different values of $h_1$.} The left $y$-axis reflects both the estimated probability that the customer accepts the offer and the normalized expected utility (to a [0,1] range). 
This figure illustrates a similar trade-off to that discussed in Section 3:
increasing the offer raises the chance that the customer will remain, but decreases the entity's expected utility. To balance these factors, we aim for the optimal expected utility, achieved at an offer of $h = 0.045$  (vertical dotted green line). This offer has an estimated acceptance probability of $0.55$ and provides an expected benefit of $4140.98$\euro{} to the supported entity.
\hfill$\triangle$

\paragraph{\textbf{Case 2. Incorporating covariates}}

Covariates $y$ may include socio-demographic information about the customer, as well as any information besides the capital he is willing to invest. These variables are typically relevant when deciding the optimal offer for each case, as we next illustrate.

Assume the setup in case 1. To account for possible changes in the customer's covariates, consider the customer declares the needed covariates to the supported entity, which conducts an aggregation process and scores him depending on the information provided. This score may be informed by the customer's socio-demographic features but could account also for extra information such as his earlier interaction with the bank and credit score history. To simplify matters, we focus on two classes of customers, say with  \textit{high} or \textit{low} score, modeled as a binary variable describing disjoint groups of potential customers.
We assume that a high-score customer will get, on average, better offers from the competitor than a low-score one. To model this, we keep the same returns offered by the competitor as in case 1 but modify the probability of the offer made: if the customer has a high score, the corresponding probabilities for the competitor's offers will be $\{ 0.025, 0.025, 0.05, 0.05, 0.05, 0.10, 0.15, 0.2, 0.3, 0.05 \}$ (\emph{i.e.} it will receive higher offers due to his more appealing profile for the entity) whereas if the customer's score is low, the probabilities will be $\{ 0.3, 0.2, 0.15, 0.10, 0.05, 0.05, $  $ 0.05, 0.05, 0.025, 0.025 \}$ (\emph{i.e.} they will reflect the weaker offers made to a customer with a less appealing profile).


\begin{figure}[htb]
\centering
\includegraphics[width=\linewidth]{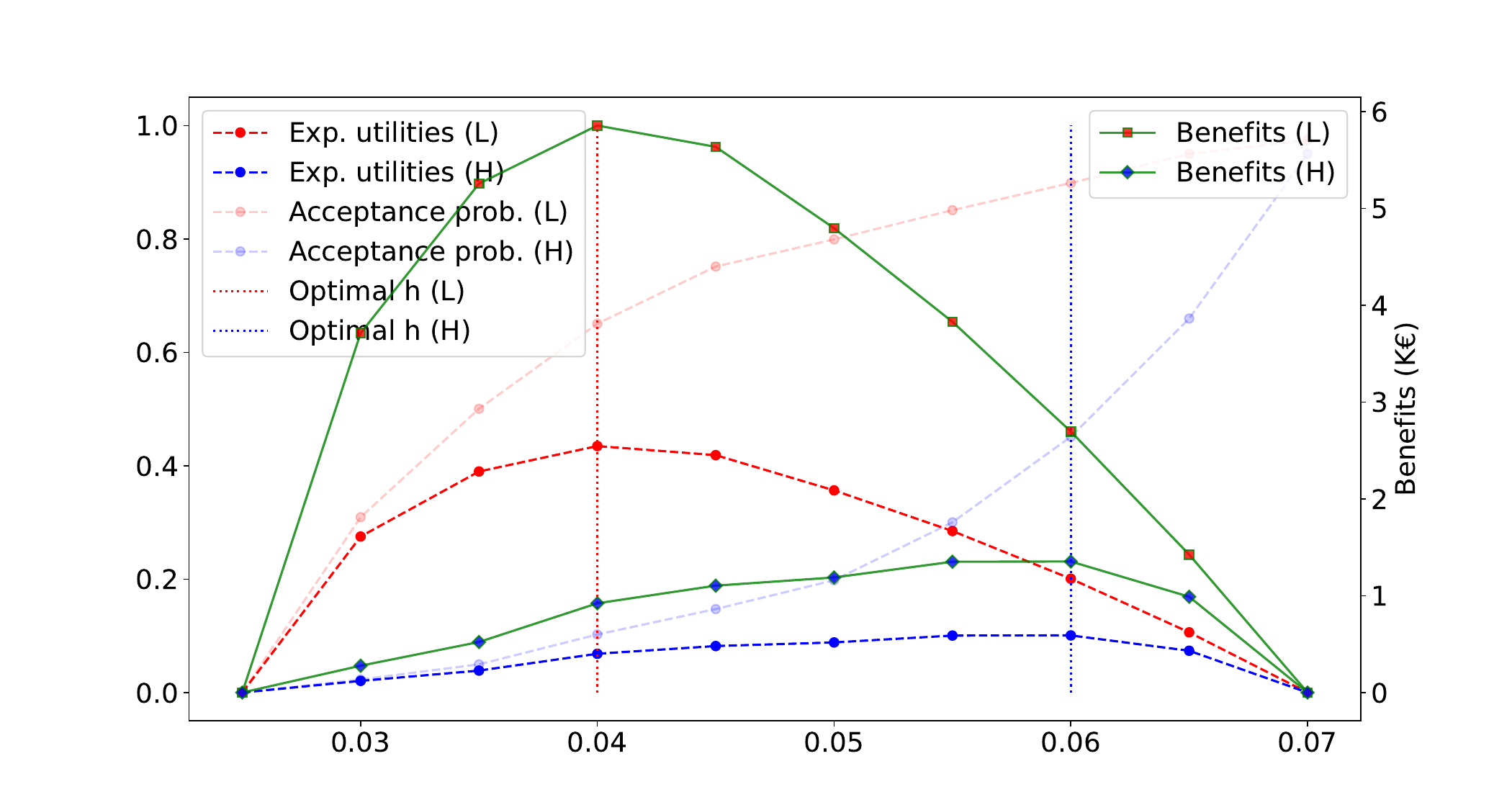}
\caption{Results for $h_1 \in [0.025, 0.065]$ accounting for two types 
 of customers: low (L), and high (H) score. Left y-axis depicts estimated 
 acceptance probabilities by customers and standardized expected utilities (best viewed in color).
}.\label{fig:static_plot_covariates}
\end{figure}

   Figure \ref{fig:static_plot_covariates} shows results comparing a low-score customer (red, L) with a high-score one (blue, H). For each of them, we present the probability of accepting our offer (faint dashed lines with dots) and the expected utility attained by the bank (solid dashed lines and dots). The optimal offer that the bank should present to each customer is shown through vertical dashed lines. Here, the left $y$-axis is 
interpreted as in Figure \ref{fig:static_plot}.
Finally, we also represent the bank's expected benefits for high-score (\textit{green line, diamonds}) and low-score (\textit{green line, squares}) customers. The expected utilities show that the optimal offers for the low and high-score customers are $h_1^* = 0.04$ (reaching almost $6.000$\euro{} in expected benefits), and $h_1 ^*=0.05$ (with approximately $1.200$\euro{} in expected benefits), respectively. 
This happens because customers with a high score tend to receive better offers from competitors. Thus, the supported bank must present higher offers to these customers. However, offering higher amounts to them does not always increase the bank's benefits, because the probability that they will accept lower offers decreases. Note that, for example, the optimal offer for these customers has only a 0.35 chance of being accepted.
In contrast, low-score customers are more likely to accept lower offers, with a 0.65 chance of accepting the optimal one. Thus, offering lower amounts to these customers is more beneficial to the bank, and the expected benefits are much higher. \hfill$\triangle$


\paragraph{\textbf{Case 3. Multiple competitors}}
 Assume now the same features for the organization we had before, but consider $n-1$ identical competitors as in case 1, with the same preference features for the potential customer.

\begin{figure}[htb]
\centering
\includegraphics[width=0.32\linewidth]{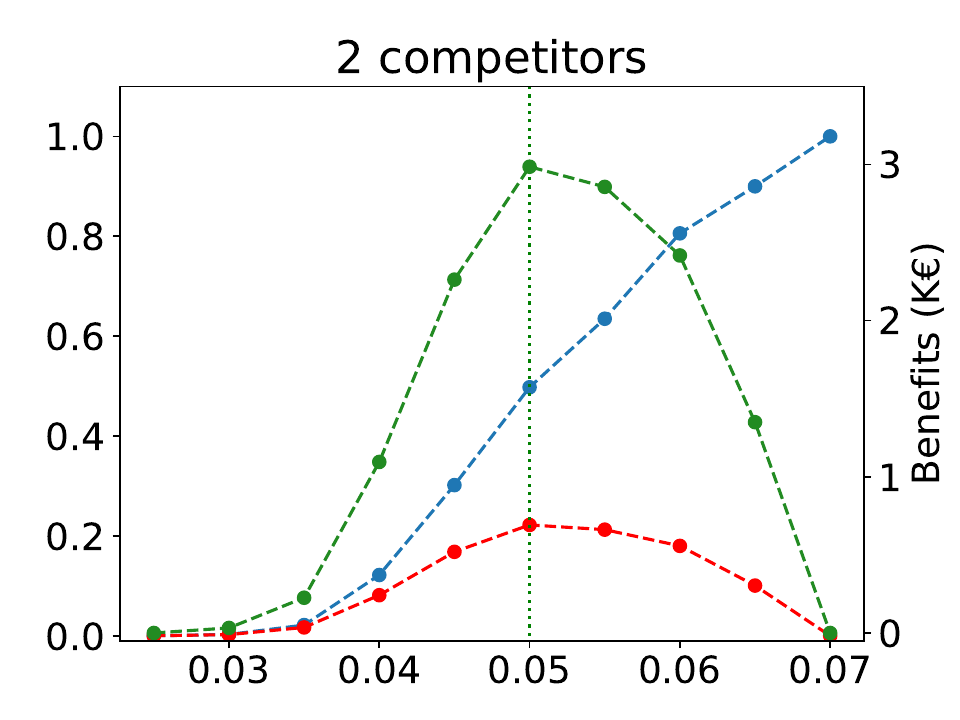}
\includegraphics[width=0.32\linewidth]{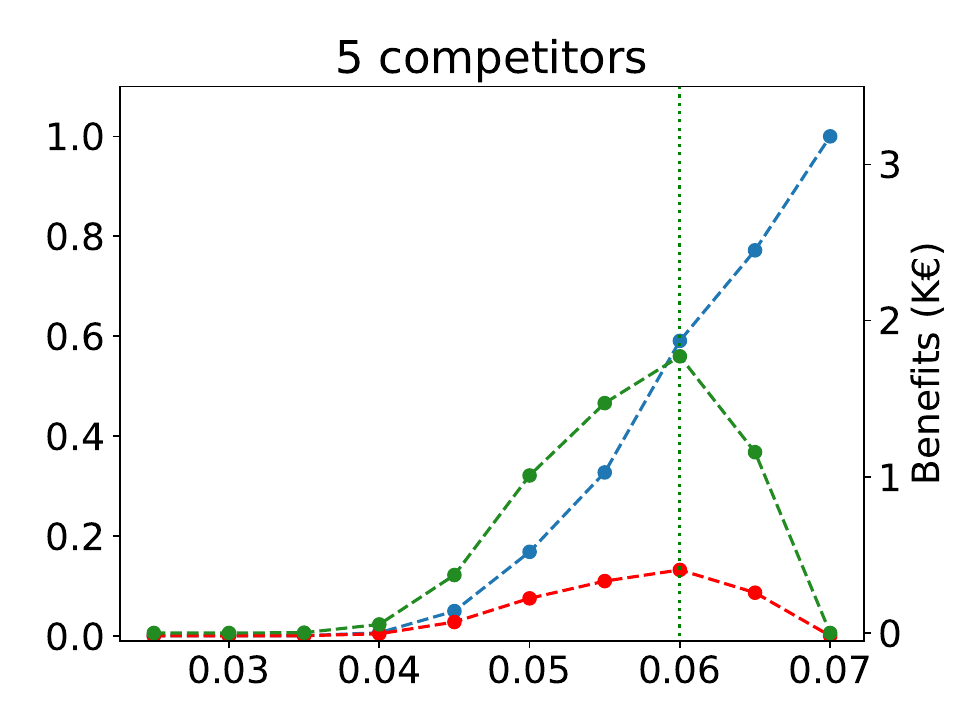}
\includegraphics[width=0.32\linewidth]{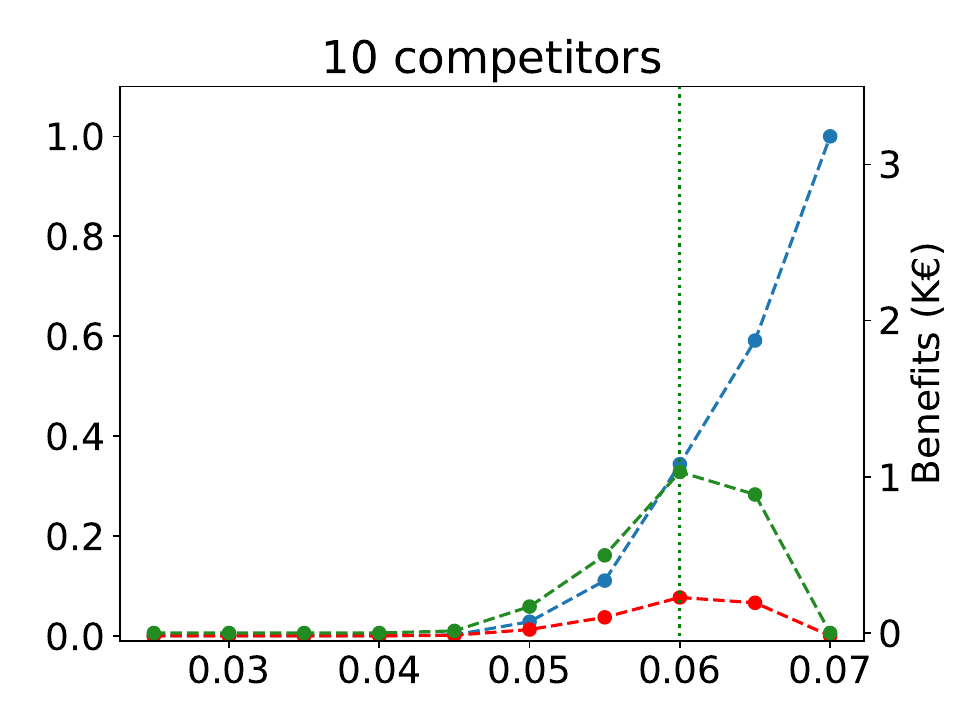}
\caption{Results depending on number of competitors for $h_1 \in [0.025, 0.065]$. 
 Left $y$-axis represents both estimated acceptance probabilities and standardized expected utilities  (best viewed in color).
}\label{fig:static_plotn}
\end{figure}

Figure \ref{fig:static_plotn} depicts the solution in the multiple competitors' setup following the convention in Figure \ref{fig:static_plot}. 
For different values of $h$, we plot the estimated probability that the customer accepts the offer (blue) and the expected utility (red) for $n=2,5,10$ competitors (\textit{left}, \textit{center} and \textit{right} figures, respectively). 

\noindent Plots illustrate the same trade-off from Figure \ref{fig:static_plot}.
   The number of competitors has a crucial impact on the results: for a given offer by the supported bank, as more competitors are considered, the probability of the customer accepting the offer decreases. As the gains of the supported bank do not change when increasing the number of competitors, the bank must increase the offer to maximize the expected utility, as the plots depict. This behavior was hinted at when discussing \eqref{eq:n_equal_competitors}. Table \ref{tab:multiple_competitors_results_pension_funds} summarizes the results as a function of the number of competitors. \hfill$\triangle$

\begin{table}[htb]
    \centering
    \caption{Results as a function of number of competitors}
    \begin{tabular}{ c  c  c  c  c }
    \toprule
    \textbf{No.competitors} & 1 & 2 & 5 & 10  \\  
    \midrule
    \textbf{Optimal offer} & 4.5 &  5    & 6  & 6  \\ \hline
    \textbf{Acc. prob.} & 0.55 &  0.49 & 0.59 & 0.34 \\ \hline
    \textbf{Exp. Utility} & 0.31 &  0.22 & 0.13  & 0.08   \\ \hline
    \textbf{Benefit (\euro{})} & 4140.98 &  2985.6 & 1771.8  & 1031.8   \\ 
    \bottomrule
    \end{tabular}
    \label{tab:multiple_competitors_results_pension_funds}
\end{table}

\section{Discussion}\label{SEC:discussion}
We have presented a framework for personalized pricing. It is based on a principled way of forecasting adversarial decisions, acknowledging business uncertainties, and promoting structured thinking about the competitors' and consumers' problems, thus enriching the solution process. 
It provides a coherent approach to competition modeling, mitigating common knowledge assumptions typical of earlier game theoretic approximations in the 
pricing domain {\color{black} \citep{Rao1972, Mesak1979, Gupta2021}}. {\color{black} Our method is designed not to compete with machine learning models but rather to complement them. ML models, particularly probabilistic ones, are useful for inferring uncertainties in pricing problems based on data. In contrast, our ARA-based approach focuses on using these inferences to prescribe optimal decisions in competitive environments. Additionally, our approach enables probabilistic forecasts of competitors' decisions when data about adversaries is scarce, which is crucial in certain strategic pricing scenarios.}

We have illustrated the versatility of the approach in two cases, one in retailing, which simplifies the proposed template, and another in pension fund markets, which enriches it. However, applications abound.
For example, another potential domain is transfer pricing, where transactions are made among companies that are part of a larger parent entity~\citep{Alles1998}.  Besides pricing decisions, we could include other important variables in the analysis, \emph{e.g.}  
the perceived quality of the product, timing of the offer, and marketing expenditure. All these factors can be combined in a similar framework to develop a general approach to product launching.

As an interesting extension of our framework, we could consider the addition of \textit{speculators} to the market: agents who acquire products from companies and resell them at a higher price 
 \citep{Su2010}. 
Moreover, it would be interesting to combine the proposed decision-making framework with statistical or ML models that estimate both customers' and competitors' behaviors when data about them is available. This could reduce uncertainty about their decisions. For example, in the pension fund market problem, if data on early customer withdrawals from pension products and relevant customer covariates are available, statistical models can be fitted to estimate their permanence and integrated into the decision-making framework. 

While the focus here was on a single customer characterized through his utility and, possibly, covariates, the approach is flexible enough to adapt to multiple customers in segmented markets, considering binomial buying processes. Future developments will also explore dynamic and adaptive strategies, extending beyond the static pricing model studied in this paper to accommodate pricing policies over time and evolving interactions.
\textcolor{black}{ In particular, we shall combine dynamic machine learning methods that forecast demand and competitors' prices time series, using appropriate covariates, with our adversarial risk analysis approach to refine these forecasts.}

\section*{Acknowledgments}
Research supported by the AXA-ICMAT Chair and the Spanish Ministry of Science program PID2021-124662OB-I00, the Severo Ochoa Excellence Programme CEX-2023-001347-S 
and a grant from the FBBVA (Amalfi). RN acknowledges the support of CUNEF Universidad. \textcolor{black}{ Discussions
 with referees are gratefully acknowledged.}

 

{\color{black}
\newpage
\appendix
\section{Appendix A}\label{App:lemmas}

This appendix provides results supporting the correct definition of 
the procedures in Sections 2.1, 2.2, and 2.3, under standard and mild assumptions.
 
We first analyse (Section 2.1) the existence of 
an optimal price $p_1^*$ for the first producer under compactness
of the feasible set of prices and continuity under the integral sign conditions. 
 
\begin{lemma}
\label{lemma_1}
If the utility function $u_1$ is continuous in $p_1$ for fixed $s$ and $c$, ${\cal P}_1$ is compact, and there exists an integrable function $\xi(c, s)$ such that $\vert u_1(p_1, c, s)\vert \leq \xi(c, s)$, then $p_1^*$ exists.
\end{lemma}
{\bf Proof.}
We follow standard ID reductions \citep{shachter} and assess at each
stage the required continuity properties.
\begin{enumerate}
    \item Eliminate node $S$, with the value node inheriting
    node $C$ as predecessor. For this, we compute the 
    expected utility with respect to $s$, given by $\psi _1 (p_1,c)= \int u_1 (p_1,c,s) q_1 (s\mid c ) ds$.
    Following the dominated convergence theorem (DCT), $\psi_1 (p_1, c)$ 
    is continuous in $p_1$ (for fixed $c$) and bounded from above
    by $\xi (c)= \int \xi(c, s) q_1 (s\mid c ) ds$.
    \item Eliminate node $C$, with the value node inheriting
    the nodes $P_2, \dots, P_n$ as predecessors.
    This leads to computing the expected utility with respect to $c$, $\psi_1 (p_1\mid p_2\,,...\,,p_n)=\sum _{c=1}^n
    \psi _1 (p_1,c)q_1 (c \mid p_1,p_2\,,...\,,p_n)$, 
    which is continuous in $p_1$ given the other prices, 
    being a convex sum of functions continuous in $p_1$, given the other $p_i$, and is dominated by $\xi (p_2\,,...\,,p_n)=
    \sum_{c=1}^n \xi(c) q_1 (c \mid p_1,p_2\,,...\,,p_n)$.
    \item Eliminate nodes $P_2, \dots, P_n$, by computing the expected utility, given by $\psi_1(p_1)= \int \dots \int  \psi_1 (p_1\mid p_2\,,\dots\,,p_n) q_1(p_2\,,\dots\,,p_n) \dd p_2 ... \dd p_n$. Again, this is continuous in $p_1$ by the DCT.
\end{enumerate}
Together with the compactness of  ${\cal P}_1$, this guarantees the existence of an optimal $p_1^*$.\hfill $\triangle$ 
\vspace{0.1in}

\noindent The conditions demanded are quite standard and easy to verify. For example, in Section 3, prices will 
typically satisfy $p_1 \in [v_1,p_1^u]$ where $v_1$ is the 
production cost and $p_1^u$ is a reasonable maximum price,
hence satisfying the compactness requirement. 
Concerning the continuity and bounds for 
$u_1$, since there is no state $s$ is in this 
case, we make the discussion  just in terms of
 $c$. In that sense, observe that $u_1 (p_1,c)$ is continuous in $p_1$  both 
for $c=1$ (as a linear function in $p_1$) and $c\neq 1$ 
(as a constant function). 
Besides, we have that $\mid u_1 (p_1, c)\mid  \leq p_1^u$.
 Similar, slightly more complex, 
analyses may be undertaken for the example in Section 4.

 Let us now pay attention to the strategic ingredient
$q_1 (c \mid p_1 ,p_2\,,...\,,p_n )$ from Section 2.2.
We model our uncertainty about the customer's preferences and beliefs through the random utilities $U_c (p_i, s)$ and random distributions $P_c (s\mid c=i)$. These, without loss of generality, are defined over a common probability space $(\Omega\,, {\cal A}\,, {\cal P})$ with atomic elements $\omega \in \Omega$ \citep{chung}. Then, we have the following result:
\begin{lemma}
\label{lemma_2}
If the utilities $u_c$ in the support of $U_c$ are almost surely (a.s.) integrable, problem (\ref{sidecars}-\ref{amish})  defines $q_1 (c \mid p_1 ,p_2\,,...\,,p_n )$.
 \end{lemma}
 {\bf Proof.}
If the utilities $u_c$ are \textit{a.s.} integrable, the random expected utilities, given by $ \int U_c(p_i, s) P_c(s\mid c=i) \dd s $ are \textit{a.s.} well-defined
 and finite. 
 Then, to obtain the required probabilities $q(c\mid p_1\,, \dots \,, p_n)$, we compare the random expected utilities between the value for the supported pricer ($p_1$) and the competitors' ($p_i \,, \forall i$), \emph{i.e.} 
$\int U_c(p_1, s) P_c(s\mid c=1) \dd s$ compared to  $\int U_c(p_i, s) P_c(s\mid c=i) \dd s$.
Therefore, the probability that the consumer chooses the first product ($ c = 1 $) given set prices $p_1\,, p_2\,, ...\,, p_n$ is that of
\[ 
    {\cal P} \left(\int U_c(p_1, s) \, P_c(s \, \mid  \, c = 1) \, \dd s \,\,
    \geq  
    \int U_c(p_i , s) \, P_c(s \, \mid  \, c = i) \, \dd s  \:\:\: \forall i  \right),
\] 
which is well-defined 
and coincides with that in \eqref{sidecars}-\eqref{amish}, Section \ref{sec:customer_problem}.\hfill $\triangle$ 
\vspace{0.1in}

Combining the ideas of the proofs of Lemmas \ref{lemma_1} and \ref{lemma_2} we obtain Lemma \ref{lemma_3}. This provides the existence of $P_2^*$ and serves to facilitate $q_1 (p_2)$ (and similarly for $q_1 (p_i)$ for 
$i=3\,,...\,,n$), where the random utility function $U_2$ and the random 
distribution $P_2$ are defined over a common underlying 
 probability space.
\begin{lemma}
\label{lemma_3}
If the utility functions $u_2$ 
in the support of $U_2 $ are \textcolor{black}{a.s.} continuous in $p_2$ for fixed $c$ and $s$, ${\cal P}_2$ is compact, and there exists an integrable function $\xi (c, s)$ such that $\vert U_2(p_2, c, s) \vert \leq \xi (c, s)$ a.s., the existence of 
 $P_2^*$ is guaranteed.
 \end{lemma}
{\bf Proof.} By the DCT and the a.s.\ continuity of the utility functions $u_2$ in the support of $U_2$, the continuity 
of the random expected utility $\Psi _2  (p_2)$ is guaranteed a.s. Together with the compactness of ${\cal P}_2$, this implies the existence of the random optimal $P_2^*$.  \hfill $\triangle$
\vspace{.1in}

General pointers to continuity and integrability of utility 
functions may be seen in \cite{French} and references quoted therein.}

\bibliography{itor}

\end{document}